# *Improved Adaptive Resolution Molecular Dynamics Simulation*

Iuliana Marin[1], Virgil Tudose[2], Anton Hadar[2], Nicolae Goga[1,3], Andrei Doncescu[4]

*Affiliation 1:* Faculty of Engineering in Foreign Languages
University POLITEHNICA of Bucharest, Bucharest, Romania, n.goga@rug.nl
*Affiliation 2:* Department of Strength of Materials, University POLITEHNICA of Bucharest, Bucharest, Romania
*Affiliation 3:* Molecular Dynamics Group, University of Groningen, Groningen, Netherlands
*Affiliation 4:* Laboratoire d'analyse et d'architecture des systemes, Université Paul Sabatier de Toulouse, Toulouse, France

*Abstract*—Molecular simulations allow the study of properties and interactions of molecular systems. This article presents an improved version of the Adaptive Resolution Scheme that links two systems having atomistic (also called fine-grained) and coarse-grained resolutions using a force interpolation scheme. Interactions forces are obtained based on the Hamiltonian derivation for a given molecular system. The new algorithm was implemented in GROMACS molecular dynamics software package and tested on a butane system. The MARTINI coarse-grained force field is applied between the coarse-grained particles of the butane system. The molecular dynamics package GROMACS and the Message Passing Interface allow the simulation of such a system in a reasonable amount of time.

## I. INTRODUCTION

The traditional computational molecular modeling indicates the general process of describing complex chemical systems in terms of a realistic atomic model, with the goal of understanding and predicting macroscopic properties based on detailed knowledge at an atomic scale [1]. In literature, the study of a system at atomistic level can be defined as fine-grained (FG) modeling [2].

Coarse-graining is a systematic way of reducing the number of degrees of freedom for a system of interest. Typically whole groups of atoms are represented by single beads and the coarse-grained (CG) force fields describe their effective interactions. Coarse-grained models are designed to reproduce certain properties of a reference system. This can be either a full atomistic model or a set of experimental data. The coarse-grained potentials are state dependent and should be re-parameterized depending on the system of interest and the simulation conditions [1].

The main disadvantage of a coarse-grained model is that the precise atomistic details are lost. In many applications it is important to preserve atomistic details for some region of special interest [2]. To combine the two systems, fine-grained and coarse-grained, it was developed a so-called multiscale simulation technique [3, 4, 5, 6, 7]. This method combines the two resolutions by coupling them with a mixing parameter $\lambda$. The multiscale interaction forces are computed as a weighted $\lambda$ sum of the interactions on fine-grained and coarse-grained levels [2].

The adaptive resolution scheme (AdResS) couples two systems with different resolutions by a force interpolation scheme. The two resolutions are called atomistic and coarse-grained [3, 4]. The atomistic representation of system describes the phenomena produced at the atomic level (noted fine-grained or FG) and the other one, coarse-grained, describes the system at molecular level (noted coarse-grained or CG).

The AdResS scheme works in the following conditions:

- the forces between molecules are scaled and the potential energy is not scaled. Consequently, the Hamiltonian cannot be used to obtain the energies;

- just non-bonded interactions between molecules are scaled. In this way, some computational resources are saved;

- particles are kept together via bonded atomistic forces that are computed everywhere, including the coarse-grained part (no pure coarse-grained in that part);

- in the transition regions there is a potential which is added (that takes into account the difference in chemical potentials between the two representations). In this paper it is denoted by $V_t$.

In the current paper, it is proposed a method for obtaining the interaction forces by applying the Hamiltonian derivation, where potentials can be scaled. The advantage of this method is that, derivation is based on the sound Hamiltonian model and energies can be reported correctly. The improved algorithm was implemented in GROMACS, a molecular dynamics software, that runs in parallel using the Message Passing Interface (MPI). Simulations have been done on butane. The MARTINI coarse-grained force field which was implemented by the Molecular Dynamics Group from the University of Groningen was also applied.

The paper is organized as follows. In the next section the improved AdResS multiscaling resolution scheme is presented. Section 3 describes the implementation of AdResS in GROMACS using MPI and its testing on a butane molecular system. Section 4 outlines the obtained results. The last section presents the final conclusions.

## II. RELATION TO EXISTING THEORIES AND WORK

In this section it is presented the relevant theory for the algorithm implemented in GROMACS – the improved AdResS multiscaling resolution scheme [3,4].

A convention is made: for a coarse-grained molecule are used capital letters for coordinates and for a fine-grained molecule, small letters. For simplifying the presentation, it is considered that the space factor $\lambda$ is function only on the $X$ coordinate of the reference system. On the $Y$ and $Z$ coordinate axis, $\lambda$ is constant.

It is denoted by $x_{ik}$ the $X$ coordinate of a fine-grained particle with the number $k$ in the coarse-grained molecule with number $i$. The coarse-grained molecule $i$ is centered in the center of mass ($X_i$) of the ensemble of corresponding fine-grained particles.

The relation between the two parameters is:

$$X_i = \frac{\sum_k m_{ik} x_{ik}}{M_i} \qquad (1)$$

$$M_i = \sum_k m_{ik} \qquad (2)$$

where $m_{ik}$ is the mass of the $k$ fine-grained particle and $M_i$ is the mass of the $i$ coarse-grained particle.

In Figure 1 more coarse-grained molecules with corresponding fine-grained particles and the parameters $x_{ik}$ and $X_i$ are represented.

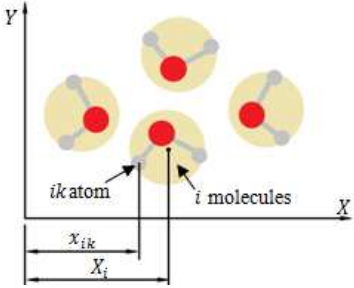

Fig. 1. *The coordinates $x_{ik}$ and $X_i$ for a coarse-grained particle*

In AdResS, the space factor $\lambda$ is function of the coordinate of a coarse-grained particle ($X_i$):

$$\lambda(X) = \cos^2\left[\frac{\sqrt{(X_i - X_m)^2} - h}{L} \cdot \frac{\pi}{2}\right] \qquad (3)$$

where $X_m$ is the coordinate in the reference system for the center of fine-grained region, $h$ is half of the length of the fine-grained region and $L$ is the length of transition (hybrid fine-grained - coarse-grained) region. In Figure 2 are displayed these parameters.

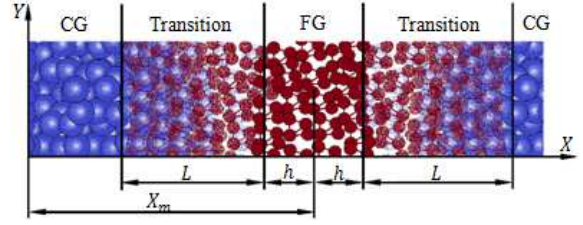

Fig. 2. *The coordinates $x_{ik}$ and $X_i$ for a coarse-grained particle*

### A. Multiscaling force

Starting from the Hamiltonian that scales the non-bonded forces, it is added an extra potential $V_t$ in the transition region that adjusts for the difference in chemical potentials between the fine-grained and coarse-grained regions.

As explained above, fine-grained bonded interactions are computed everywhere, including the coarse-grained region. The potential energy in the multiscaling model ($V^m$) is a function of coordinates only and it is chosen as: The multiscaling factor is defined as follows:

$$V^m(x) = \lambda(X(x)) \cdot V_{FG}^{nb}(x) + [1 - \lambda(X(x))] \cdot V_{CG}^{nb}(X(x)) + V_{FG}^b(x) + V_{CG}^b(X(x)) + V_t \qquad (4)$$

The multiscaling factor is defined as follows:

$$\lambda(X) = \cos^2\left[\frac{\sqrt{(X_i - X_m)^2} - h}{L} \cdot \frac{\pi}{2}\right] \qquad (5)$$

Using (4) in (5), it is obtained:

$$-F^m(x_{ik}) = \frac{\partial \lambda}{\partial x_{ik}} \cdot V_{FG}^{nb} + \lambda \cdot \frac{\partial V_{FG}^{nb}}{\partial x_{ik}} + (1 - \lambda) \cdot \frac{\partial V_{CG}^{nb}}{\partial x_{ik}} - \frac{\partial \lambda}{\partial x_{ik}} \cdot V_{CG}^{nb} + \frac{\partial V_{FG}^b}{\partial x_{ik}} + \frac{\partial V_{CG}^b}{\partial x_{ik}} + \frac{\partial V_t}{\partial x_{ik}} \qquad (6)$$

Because $\lambda$ is function of $X_i$, it is necessary to change the variable. Then

$$\frac{\partial \lambda}{\partial x_{ik}} = \frac{\partial \lambda}{\partial X_i} \cdot \frac{\partial X_i}{\partial x_{ik}} \qquad (7)$$

According to (1), it is obtained

$$\frac{\partial X_i}{\partial x_{ik}} = \frac{m_{ik}}{M_i} \qquad (8)$$

Using (8) in (7), results

$$\frac{\partial \lambda}{\partial x_{ik}} = \frac{m_{ik}}{M_i} \cdot \lambda' \qquad (9)$$

where $\lambda'$ is the derivative of (3) with respect $X$ which is computed at the end of section (see formula (18)).

Taking that into account

$$\frac{\partial V_{FG}^{nb}}{\partial x_{ik}} = -F_{ik}^{nb} \quad (10)$$

$$\frac{\partial V_{FG}^{b}}{\partial x_{ik}} = -F_{ik}^{b} \quad (11)$$

are the non-bonded, respectively bonded, forces in the atom $ik$, and

$$\frac{\partial V_{CG}^{nb}}{\partial X_i} = -F_i^{nb} \quad (12)$$

$$\frac{\partial V_{CG}^{b}}{\partial X_i} = -F_i^{b} \quad (13)$$

are the non-bonded, respectively bonded, force in the coarse-grained molecule $i$, formula (6) becomes:

$$-F^m(x_{ik}) = \frac{m_{ik}}{M_i} \cdot \lambda' \cdot V_{FG}^{nb} - \lambda \cdot F_{ik}^{nb} + (1-\lambda) \cdot \frac{\partial V_{CG}^{nb}}{\partial X_i} \cdot \frac{\partial X_i}{\partial x_{ik}} - \frac{m_{ik}}{M_i} \cdot \lambda' \cdot V_{CG}^{nb} - F_{ik}^{b} + \frac{\partial V_{CG}^{b}}{\partial X_i} \cdot \frac{\partial X_i}{\partial x_{ik}} + \frac{\partial V_t}{\partial x_{ik}} \quad (14)$$

By making the notation

$$\frac{\partial V_t}{\partial x_{ik}} = -F_t \quad (15)$$

it is obtained:

$$-F^m(x_{ik}) = \frac{m_{ik}}{M_i} \cdot \lambda' \cdot V_{FG}^{nb} - \lambda \cdot F_{ik}^{nb} - (1-\lambda) \cdot F_i^{nb} \cdot \frac{m_{ik}}{M_i} - \frac{m_{ik}}{M_i} \cdot \lambda' \cdot V_{CG}^{nb} - F_{ik}^{b} - F_i^{b} \cdot \frac{m_{ik}}{M_i} - F_t \quad (16)$$

Then it results:

$$F^m(x_{ik}) = -\frac{m_{ik}}{M_i} \cdot \lambda' \cdot V_{FG}^{nb} + \lambda \cdot F_{ik}^{nb} + F_{ik}^{b} + \frac{m_{ik}}{M_i} \cdot \lambda' \cdot V_{CG}^{nb} + (1-\lambda) \cdot F_i^{nb} \cdot \frac{m_{ik}}{M_i} + F_i^{b} \cdot \frac{m_{ik}}{M_i} + F_t \Rightarrow$$

$$F^m(x_{ik}) = \left(-\frac{m_{ik}}{M_i} \cdot \lambda' \cdot V_{FG}^{nb} + \lambda \cdot F_{ik}^{nb} + F_{ik}^{b}\right) + \frac{m_{ik}}{M_i}\left[\lambda' \cdot V_{CG}^{nb} + (1-\lambda) \cdot F_i^{nb} + F_i^{b}\right] + F_t \quad (17)$$

Finally, the derivative of $\lambda$ with respect $X$ is obtained from formula (3):

$$\frac{\partial \lambda}{\partial X_i} = \lambda' = \frac{\pi}{L} \cdot \cos\left[\frac{\sqrt{(X_i-X_m)^2}-h}{L} \cdot \frac{\pi}{2}\right] \cdot \sin\left[\frac{\sqrt{(X_i-X_m)^2}-h}{L} \cdot \frac{\pi}{2}\right] \cdot \frac{X_m-X_i}{\sqrt{(X_i-X_m)^2}} \quad (18)$$

### III. TECHNOLOGY APPROACH

The implementation of the improved AdResS is done in GROMACS, a package specialized for running molecular dynamics simulations. GROMACS was firstly developed at the University of Groningen. Because the simulation time might last for a long time, even in terms of months, the Message Passing Interface (MPI) is used for parallelizing the computations within the molecular system. The thermostat that was used is stochastic dynamics.

The implementation of the new algorithm is based on MPI parallelization (see Figure 3). MPI is used by dividing the simulation box into several boxes. The number of processors involved in computation is equal to the number of the smaller simulation boxes. Data is passed between the neighboring simulation boxes through the use of messages which is constant. The MARTINI coarse-grained force field which is applied on the particles of the system lowers the initial energy of atoms until the reference temperature is reached and it is maintained at that state [8].

The parallelization and the improved AdResS algorithm are depicted in Figure 3.

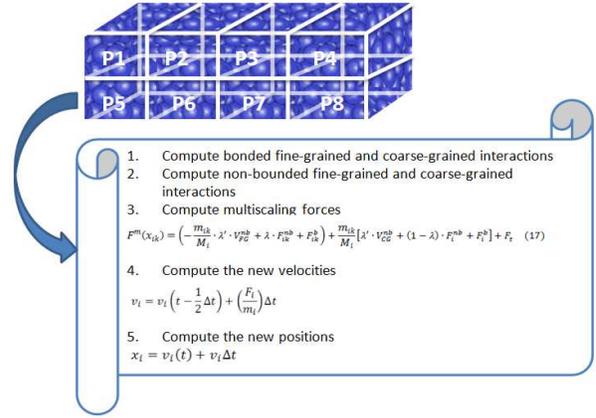

Fig. 3. *The simulation box and its processes*

Parallelization is done through MPI. The computation is divided between the involved processors. Each processor will follow the algorithm that starts by firstly computing the bonded fine-grained and coarse-grained interactions $V_{FG}^b, V_{CG}^b$. The next step is the computation of the non-bounded fine-grained and coarse-grained interactions $V_{FG}^{nb}, V_{CG}^{nb}$. The multiscaling forces are computed according to formula (17). The velocities and the positions of the particles get after that updated.

### IV. FINDINGS

The simulation was done on an Asus ZenBook Pro UX501VW computer having an Intel Core i7-6700HQ processor with 8 threads and a frequency of 2.6 GHz. The RAM has a capacity of 16 GB. The simulation box comprises butane with 36900 atoms. The dimension of an atom is equal to 10 nm on each direction (x, y, z). The reference temperature is set at 323 K. The system of atoms has been simulated on a different number of processors for 10,000 simulation steps. For each processor, a number of eight executions were made, each of the reported values being an average of the eight execution times.

The execution time expressed in ns/day and in hour/ns are presented in Table 1.

TABLE I. EXECUTION TIMES AND TEMPERATURE FOR A DIFFERENT NUMBER OF PROCESSORS

| Test No. | Performance | | Temperature |
|---|---|---|---|
| | [ns/day] | [hour/ns] | [K] |
| 1 | 7.421 | 3.234 | 326.487 |
| 2 | 15.326 | 1.566 | 326.497 |
| 3 | 20.146 | 1.191 | 326.440 |
| 4 | 26.127 | 0.919 | 326.481 |
| 5 | 19.328 | 1.242 | 326.543 |
| 6 | 28.193 | 0.851 | 326.498 |
| 7 | 23.697 | 1.013 | 326.525 |
| 8 | 35.591 | 0.674 | 326.466 |

In the table above the standard deviations are no larger than 0.036 for the execution time and 0.110 for the temperature.

A different number of processors was used to analyze how parallelization evolves at run time for the butane system. The execution time according to the number of processors is presented in Fig. 4.

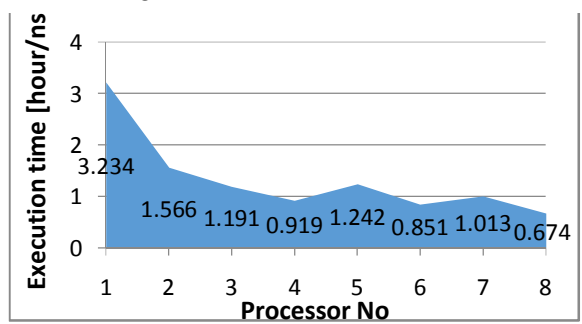

Fig. 4. *Execution time expressed in hour/ns as a function of the processor number used*

It can be observed that as the number of processors increased, the execution time linearly decreased, after which a plateau was obtained for the considered butane system with 36900 atoms.

There is a significant difference in time for the simulation where only one processor is involved compared to the case when several processors interact through the use of MPI - a speedup of about four times more was obtained when six processors were used as compared with the case of one processor.

In Figure 5 is represented the variation of temperature in time. By using the stochastic dynamics thermostat and the improved AdResS, the temperature of the system is kept around the value of 326 K as it can be observed in Table 1.

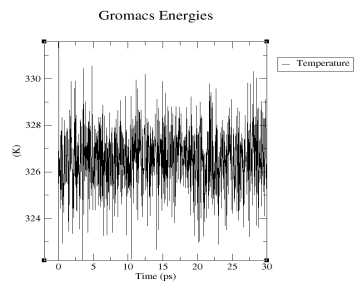

Fig. 5. *Temperature variation in time*

The reference temperature for the system was of 323 K. The obtained value with a difference of around ± 3K is within the statistical accepted errors for such simulations.

V. CONCLUSIONS

The article presents the improved AdResS through the MPI parallelization implemented in the GROMACS molecular dynamics software and its testing on a butane molecular system, along with the experimental results. Usually the simulation time is large. For this reason parallelization is used for obtaining the results in a shorter period of time.

The algorithm depends on the hardware used and on the number of processors on which it is run. There is a significant difference in time for the simulation where only one processor is involved as compared to the case when several processors interact through the use of MPI - a speedup of about four times more was obtained when six processors were used as compared with the case of one processor. The temperature is kept within the accepted statistical ranges for such simulations.

Further work include the testing of this implementation on more systems, generalizing the theory for cylinder coordinates, further improvements for AdResS algorithm.

REFERENCES

[1] M. J. Abraham, D. Van Der Spoel, E. Lindahl, B. Hess and the Gromacs Development Team, "GROMACS User Manual version 4.6.7", *www.gromacs.org*, p. 110, 2014.
[2] N. Goga et al., "Benchmark of Schemes for Multiscale Molecular Dynamics Simulations", J. Chem. Theory Comp., vol. 11, 2015, pp. 1389-1398.
[3] M. Praprotnik, L. Delle Site, K. Kremer, "Adaptive resolution molecular-dynamics simulation: Changing the degrees of freedom on the fly", J. Chem. Phys. 2005.
[4] M. Praprotnik, L. Delle Site, K. Kremer, "Multiscale simulation of soft matter: From scale bridging to adaptive resolution", Annu. Rev. Phys. Chem., vol. 59, 2008, pp.545-571.
[5] M. Christen, W. F. van Gunsteren, "Multigraining: An algorithm for simultaneous fine-grained and coarse-grained simulation of molecular systems", J. Chem. Phys., vol. 124, 2006.
[6] B. Ensing, S. O. Nielsen, P. B. Moore, M. L. Klein, M. J. Parrinello, "Energy conservation in adaptive hybrid atomistic/coarse-grain molecular dynamics", J. Chem. Theory Comp., vol. 3, 2007, pp. 1100-1105.
[7] L. Delle Site, "What is a Multiscale Problem in Molecular Dynamics?", Entropy, vol. 15, 2014, pp. 23-40.
[8] S. J. Marrink, H.J. Risselada, S. Yefimov, D. P. Tieleman, A. H. de Vries, "The MARTINI Force Field: Coarse Grained Model for Biomolecular Simulations", *J. Phys. Chem.*, 111, pp. 7812-7824, 2007.